\documentclass[aps,prl,twocolumn,superscriptaddress]{revtex4-2}
\pdfoutput=1
\usepackage[T2A,T1]{fontenc}
\makeatletter
\def\easycyrsymbol#1{\mathord{\mathchoice
  {\mbox{\fontsize\tf@size\z@\usefont{T2A}{\rmdefault}{m}{n}#1}}
  {\mbox{\fontsize\tf@size\z@\usefont{T2A}{\rmdefault}{m}{n}#1}}
  {\mbox{\fontsize\sf@size\z@\usefont{T2A}{\rmdefault}{m}{n}#1}}
  {\mbox{\fontsize\ssf@size\z@\usefont{T2A}{\rmdefault}{m}{n}#1}}
}}
\makeatother

\usepackage{xcolor}
\usepackage{slashed}
\usepackage{amsmath}  
\usepackage{amssymb}  
\usepackage{mathtools}
\usepackage{mathrsfs}
\usepackage{array}
\usepackage{rotating}
\usepackage{multirow}
\usepackage{hhline}

\begin{document}

\begin{flushright}
	DESY 22-039\\
\end{flushright}

\title{Duality Symmetric Electrodynamics in Curved Spacetime}

\author{Anton V. Sokolov}
\email[e-mail: ]{anton.sokolov@physics.ox.ac.uk}

\affiliation{Rudolf Peierls Centre for Theoretical Physics, University of Oxford, Parks Road, Oxford OX1 3PU, United Kingdom}
\affiliation{Deutsches Elektronen-Synchrotron DESY, Notkestr. 85, 22607 Hamburg, Germany}

\begin{abstract}
We derive Maxwell equations for electric and magnetic fields in curved spacetime from first principles, relaxing an unnecessary assumption on the structure of the four-potential inherent to the standard approach and thus restoring the full consistency with the equivalence principle in the following two important cases: first, if the electromagnetic field is considered as a physical entity separate from the charged particles used to measure it, and second, if hypothetical magnetically charged particles are allowed to exist.
We find that in a generic curved spacetime, the electromagnetic field has to be described by two pairs of electric and magnetic fields, instead of the only one pair which is enough in the flat spacetime case.
\end{abstract}


\maketitle

\begingroup
\allowdisplaybreaks
\textit{Introduction.}---These days, we routinely detect kHz frequency Gravitational Waves (GWs) with laser interferometers such as Advanced LIGO~\cite{LIGOScientific:2014pky}, Advanced VIRGO~\cite{VIRGO:2014yos} and KAGRA~\cite{KAGRA:2020tym}. One of the challenges ahead is to detect GWs in other types of experiments which would probe different wavelengths. A lot of work has been done in this direction. In particular, it has been recently argued that the existing axion detection experiments, or versions of them, can allow for GW detection in the Ultra-High Frequency (UHF), i.e. above a kHz, range~\cite{Ejlli:2019bqj, Ringwald:2020ist, Berlin:2021txa, Domcke:2022rgu}. To interpret the results of such experiments, one has to model the influence of the passing GWs on the magnetic and electric fields of the detector, which can be done by considering Maxwell equations in the GW background~\cite{1970NCimB..70..129B, 1971NCimL...2..549B}. These equations are obtained from the well-known formulation of classical electrodynamics in curved spacetime~\cite{Einstein1916} by A.~Einstein. As we will show, however, contrary to the flat spacetime Maxwell equations, these equations are not symmetric with respect to the following interchange of the electric and magnetic fields $\mathbf{E} \rightarrow \mathbf{B}$, $\mathbf{B} \rightarrow -\mathbf{E}$, or more generally, with respect to the $SO(2)$ electric-magnetic duality symmetry transformations, although this symmetry has a physically important conserved Noether charge associated to it. This issue leads to a number of tensions within the standard approach. In particular, as it will be discussed in the next section, whenever one aims to fully quantize the electromagnetic field with the known methods, or considers a thought experiment involving a hypothetical magnetic charge, the conventional coupling of electromagnetism to gravity violates the equivalence principle (EP).
Since no experiment to date has detected violations of the EP, we argue that the coupling of electromagnetic (EM) field to gravity fully respecting the EP is preferred over the conventionally used coupling. Therefore, we find the corresponding preferred general-covariant formulation of electrodynamics which is fully consistent with the electric-magnetic duality symmetry. Intriguingly, we find that the known set of curved spacetime Maxwell equations has to be complemented by another set of equations featuring a separate pair of electric and magnetic fields acting on magnetically charged particles.

\textit{Electric-magnetic duality symmetry in presence of gravity.}---Maxwell equations for free EM fields are:
\begingroup
\allowdisplaybreaks
\begin{eqnarray} \label{max1}
  &&\partial_{\mu} F^{\mu \nu} = 0 \, , \\[3pt] \label{max2}
  &&\epsilon^{\mu \nu \lambda \rho}\, \partial_{\nu} F_{\lambda \rho} = 0 \, ,
\end{eqnarray}
\endgroup
where $F_{\mu \nu} = \partial_{\mu} A_{\nu} - \partial_{\nu} A_{\mu}$ is the EM field strength tensor, $A_{\mu}$ is the four-potential of the EM field, $\epsilon^{\mu \nu \lambda \rho}$ is the Levi-Civita symbol, $\epsilon^{0123} = 1$.
In his seminal work on General Relativity (GR)~\cite{Einstein1916}, using the EP as a guide, A.~Einstein states that the effects gravity exhibits on matter are equivalent to the effects of curved spacetime. Adapting the Eqs.~\eqref{max1} and~\eqref{max2} to the case of curved spacetime, he obtains:
\begin{eqnarray}\label{ein1}
    && D_{\mu} F^{\mu \nu} = 0 \, , \\[3pt] \label{ein2}
    && \varepsilon^{\mu \nu \lambda \rho} D_{\nu} F_{\lambda \rho} = 0 \, ,
\end{eqnarray}
where $D_{\mu}$ is a covariant derivative, $\varepsilon^{\mu \nu \lambda \rho } =  \epsilon^{\mu \nu \lambda \rho} / \sqrt{-g}$ is the Levi-Civita pseudotensor, $g \equiv \textrm{det}\, g_{\mu \nu}$ with $g_{\mu \nu}$ being the spacetime metric tensor, for which we adopt the $(-,+,+,+)$ signature. Note that $F_{\mu \nu} = D_{\mu} A_{\nu} - D_{\nu} A_{\mu} = \partial_{\mu} A_{\nu} - \partial_{\nu} A_{\mu}$.

Defining electric and magnetic fields in terms of the covariant components of $F_{\mu \nu}$~\footnote{Note that one defines physical fields in terms of the \textit{covariant} components of the field strength tensor since the latter tensor is proportional to the commutator of $U(1)$ covariant derivatives, while in GR, derivatives are essentially defined with lower indices.}:
\begin{eqnarray}\label{covar1}
E_i = -F_{0i}\, , \;\; B_i = \frac{1}{2}\, \epsilon_{ijk} F_{jk}\, ,
\end{eqnarray}
one finds the following vector form for the Maxwell equations:
\begingroup
\allowdisplaybreaks
\begin{eqnarray}\label{nomm}
&&\pmb{\nabla}\! \cdot \! \mathbf{B} = 0 \, , \\[3pt] \label{Fara}
&& \pmb{\nabla}\! \times \! \mathbf{E} + \frac{\partial  \mathbf{B} }{\partial t} = 0 \, ,  \\[3pt] \label{nomm1}
&& \pmb{\nabla}\! \cdot \! \left\lbrace \sqrt{-g} \left[ g^{00} \mathcal{G}\!\cdot \! \mathbf{E} - \mathbf{g} \left( \mathbf{g}\! \cdot \! \mathbf{E} \right) + \mathcal{G} \! \cdot \! \left( \mathbf{g}\! \times \! \mathbf{B} \right) \right] \right\rbrace = 0 \, , \;\; \\[3pt]
&& \pmb{\nabla}\! \times \! \left\lbrace \sqrt{-g} \left[ \mathbf{g} \!\times \! \left( \mathcal{G}\! \cdot \! \mathbf{E} \right)  + \frac{1}{2}\, \mathrm{tr} \left( \mathcal{G} \mathbf{S} \mathcal{G} \mathbf{S} \! \cdot \! \mathbf{B} \right)  \right] \right\rbrace - \nonumber \\ \label{mmlas}
&&\frac{\partial}{\partial t} \left\lbrace \sqrt{-g} \left[ g^{00} \mathcal{G}\!\cdot \! \mathbf{E} - \mathbf{g} \left( \mathbf{g}\! \cdot \! \mathbf{E} \right) + \mathcal{G} \! \cdot \! \left( \mathbf{g}\! \times \! \mathbf{B} \right) \right] \right\rbrace = 0 \, , \;\;\;\;\;
\end{eqnarray}
\endgroup
where $\mathcal{G} \equiv g^{ij}$, $g_i \equiv g^{0i}$, $(S_i)_{jk} \equiv \epsilon_{ijk}$, Latin letters denote spatial indices~\cite{1994AmJPh..62..923C}. Note that the first two vector equations~\eqref{nomm} and \eqref{Fara} coincide with the ordinary expressions for the no magnetic monopoles condition and the Faraday law, respectively. 

One can easily check that the Maxwell equations~\eqref{nomm}--\eqref{mmlas} are generally not invariant with respect to the $SO(2)$ electric-magnetic duality transformations:
\begin{equation}
\begin{aligned}\label{rot}
    &\mathbf{E} \; \rightarrow \; \mathbf{E} \cos \theta + \mathbf{B} \sin \theta \\
    &\mathbf{B} \; \rightarrow \; \mathbf{B} \cos \theta - \mathbf{E} \sin \theta \, \, ,
\end{aligned}
\end{equation}
which however leave the flat spacetime Maxwell equations unchanged. In flat spacetime, rotations~\eqref{rot} are known to be induced by the following infinitesimal transformation of the transverse part of the vector-potential $\mathbf{A}^{\text{T}}$~\cite{Deser:1976iy}:
\begin{equation}
    \delta \mathbf{A}^{\text{T}} = -\theta \, \pmb{\nabla}^{-2}\, \pmb{\nabla} \! \times \! \dot{\mathbf{A}}^{\text{T}} \, ,
\end{equation}
which leaves the EM field action invariant. The latter invariance gives rise to the conserved $U(1)$ charge:
\begin{eqnarray}\label{helic}
    &&S_0 =\int d^3 x\, s_0 = \nonumber \\
    && \frac{1}{2} \int d^3 x \left\lbrace \mathbf{A}^{\text{T}} \! \cdot \pmb{\nabla}\!\times \! \mathbf{A}^{\text{T}} - \dot{\mathbf{A}}^{\text{T}} \! \cdot \pmb{\nabla}^{-2}\, \pmb{\nabla} \! \times \! \dot{\mathbf{A}}^{\text{T}} \right\rbrace = \nonumber \\
    && -\frac{1}{2} \int d^3 x \Bigl\lbrace \mathbf{B} \cdot \pmb{\nabla}^{-2}\, \pmb{\nabla} \! \times \! \mathbf{B} + \mathbf{E}  \cdot \pmb{\nabla}^{-2}\, \pmb{\nabla} \! \times \! \mathbf{E} \Bigr\rbrace \, , \qquad
\end{eqnarray}
which is known as helicity of EM field~\cite{Zwanziger:1968rs, Deser:1976iy, Afanasev:1995vh, Trueba_1996, Cameron_2012, Bliokh:2012zr}. The two terms in Eq.~\eqref{helic} have similar bilinear structure in terms of $\mathbf{E}$ and $\mathbf{B}$ and their time derivatives cancel each other due to the similarity of the structure of the Maxwell equations~\eqref{max1} and~\eqref{max2} for $\dot{\mathbf{E}}$ and $-\dot{\mathbf{B}}$. Introducing the dual-symmetrized spin angular momentum density of EM field $\mathbf{s}$, one can rewrite the conservation law in terms of the conservation of the four-current $\partial_{\mu } s^{\mu} = 0$. Adapting this law to the case of curved spacetime, one obtains: 
\vspace*{-2pt}
\begin{equation}\label{cons}
D_{\mu} s^{\mu} = 0 \, ,
\vspace*{-2pt}
\end{equation}
i.e. the current $s^{\mu}$ corresponding to the internal $U(1)$ symmetry of the EM field~\eqref{rot} has to be covariantly conserved. However, it is easy to see that the latter conservation is inconsistent with Eqs.~\eqref{ein1} and~\eqref{ein2} (i.e. Eqs.~\eqref{nomm}-\eqref{mmlas}), since they explicitly break this $U(1)$ symmetry~\footnote{It was argued in Ref.~\cite{Deser:1976iy} that, contrary to our conclusion, conventional electrodynamics in curved spacetime does preserve the electric-magnetic duality symmetry; however a careful investigation shows that their calculation is incorrect.}.

Thus, Eqs.~\eqref{ein1}, \eqref{ein2} and~\eqref{cons} cannot be satisfied simultaneously. This suggests that the conventional coupling of electromagnetism to gravity introduced in Ref.~\cite{Einstein1916} is \textit{not} minimal, i.e. its effects on the EM field are different from what one would get by simply considering classical EM field in a curved spacetime background. In this formalism, gravity violates the internal symmetry of the EM field: it interacts with distinct helicity states of the EM field differently. If the EM field is regarded as a physical entity in itself, separate in principle from the charged particles used to probe it, the breaking of its internal symmetry by gravity may be regarded as a violation of the EP and so signify a conceptual problem within the standard formalism. While all exhaustive quantum-field-theoretic approaches to EM interactions are indeed based on the perspective that the EM field should be quantized separately from the particles used to probe it, an alternative point of view on the electromagnetic field is in principle possible in the classical theory, see for example Ref.~\cite{Wheeler:1945ps}. It is interesting that if one adopts this alternative point of view on the EM field, there is no immediate inconsistency anymore.
Indeed, if an observer can use only electric charges to measure the EM field, the measurement process itself breaks the electric-magnetic duality symmetry~\footnote{Note that the breaking of the electric-magnetic duality symmetry by sources of the EM field is of no relevance to the present discussion, because the coupling of the EM field to gravity cannot depend on the kind of charges sourcing the EM field: in fact, these charges could be causally disconnected from a given gravitational field.}: the symmetry is actually never present. The full quantization of the theory in such approach to EM interactions is however an open problem.

The tension with the EP becomes even more clear if one considers a thought experiment involving a hypothetical magnetic charge. For example, consider a pair of probe charges of the same mass initially at rest, one of which is electric and another one is magnetic, in the background of static parallel electric and magnetic fields of the same magnitude. The system obviously respects the electric-magnetic duality symmetry. Suppose now one switches on a generic gravitational field. According to the Eqs.~\eqref{ein1} and~\eqref{ein2} (i.e. Eqs.~\eqref{nomm}-\eqref{mmlas}), the electric and magnetic fields acting on the probe charges change differently. By registering the difference in motion of the charges, one can clearly state that it was the effect of gravity, but not of the acceleration of the laboratory; the EP is violated.

\textit{Two-potential formulation of electrodynamics in curved spacetime.}---From the results of the previous section, it may seem that the minimal coupling of EM field to gravity does not exist. In fact, this is not true, since the inconsistency with the EP found in the previous section can be traced back to a certain assumption made in Ref.~\cite{Einstein1916}. Indeed, consider a generic experiment where one aims to probe the coupling of EM field to gravity. An experimenter creates probe EM fields in the laboratory and measures their change in a varying background gravitational field, such as the one of a GW. While the probe EM fields are created by electric currents and thus can be described by a \textit{regular} four-potential, the same cannot be a priori stated for gravity-induced variations of these fields. It is well-known that some physical configurations of EM field, e.g. field of a magnetic monopole~\cite{Dirac:1931kp}, cannot be described by a regular four-potential. This means that one should not a priori restrict the solution space for the gravity-induced EM four-potential by solely regular configurations. However, this is exactly what the Bianchi identity~\eqref{ein2} does:
\begin{equation}
    \varepsilon^{\mu \nu \lambda \rho} D_{\nu} F_{\lambda \rho} = 0 \quad \Rightarrow \quad \left[ \partial_{\nu}, \partial_{\lambda} \right] A_{\rho} = 0 \, ,
\end{equation}
where the latter expression follows from the former one regardless of the metric. Such unphysical restriction on the solution space is the reason for the inconsistency discussed in the previous section.

Not to miss any possible physical configurations of the gravity-induced EM fields, one can either allow four-potential $A_{\mu}$ to be non-regular, so that $\left[ \partial_{\nu}, \partial_{\lambda} \right] A_{\rho}$ need not be zero everywhere, or introduce an additional four-potential $C_{\mu}$ and work in the two-potential formulation of electrodynamics~\cite{Schwinger:1966nj, Zwanziger:1968rs}. For the sake of simplicity, let us choose the second option, in particular we will build on the two-potential formulation by Zwanziger~\cite{Zwanziger:1968rs}. We introduce the following general-covariant Lagrangian density:
\begin{eqnarray}\label{Lgr}
   \mathcal{L}_1 =  -\frac{1}{2}\, n^{\alpha'}\! n^{\gamma'}\! e^{\alpha}_{\alpha'} e^{\gamma}_{\gamma'} \sqrt{-g}\, \Biggl\lbrace \frac{1}{2}\, g_{\zeta \gamma} \varepsilon^{\zeta \beta \sigma \rho} \Bigl( E_{\alpha \beta} G_{\sigma \rho} - && \Bigr. \Biggr. \nonumber \\
    \Biggl. \Bigl. G_{\alpha \beta} E_{\sigma \rho} \Bigr) +  g^{\rho \beta} \Bigl( E_{\alpha \beta} E_{\gamma \rho} + G_{\alpha \beta} G_{\gamma \rho} \Bigr) \Biggr\rbrace \, , && \quad
\end{eqnarray}
where $e^{\alpha}_{\alpha'}$ are vierbein vectors and $n^{\alpha'}\!$ is a fixed Lorentz four-vector with unit norm $n^2 = 1$; we denote Lorentz indices by Greek letters with a prime. The dynamical variables of the theory are the two four-potentials $A_{\mu}$ and $C_{\mu}$, which enter the Lagrangian density~\eqref{Lgr} via the gauge-invariant field strength tensors:
\begin{eqnarray}\label{fieldstr}
    &&E_{\alpha \beta} = D_{\alpha} A_{\beta} - D_{\beta} A_{\alpha} = \partial_{\alpha} A_{\beta} - \partial_{\beta} A_{\alpha} \, , \\
    && G_{\alpha \beta} = D_{\alpha} C_{\beta} - D_{\beta} C_{\alpha} = \partial_{\alpha} C_{\beta} - \partial_{\beta} C_{\alpha} \, .
\end{eqnarray}
In flat spacetime $e^{\alpha}_{\alpha'} = \delta^{\alpha}_{\alpha'}$, the Lagrangian density~\eqref{Lgr} reduces to the Lagrangian density introduced in Ref.~\cite{Zwanziger:1968rs}. Similarly to the case of the latter article, a very interesting feature of the Lagrangian density~\eqref{Lgr} is that while it contains a fixed four-vector $n^{\alpha'}\!$ and is therefore not manifestly local-Lorentz-invariant, the motion of the system is fully local-Lorentz-invariant~\footnote{It is even more remarkable that the Lorentz-invariance in the flat spacetime case is known to hold also at the quantum level: the partition function of the Zwanziger theory was shown to be Lorentz-invariant in Refs.~\cite{Brandt:1977be, Brandt:1978wc}}. We will illustrate that this property indeed holds for the general-covariant case shortly after deriving the equations of motion (EOM).
\endgroup

Varying the action $S_1 = \int d^4 x\, \mathcal{L}_1$ with respect to the four-potentials $A_{\mu}$ and $C_{\mu}$, we obtain the following EOM for the electromagnetic field:
\begin{eqnarray}\label{eom1}
    && \partial_{\zeta} \Bigl( \sqrt{-g}\, g^{\rho \beta} \gamma^{\zeta} \gamma^{\alpha} E_{\alpha \beta} - \sqrt{-g}\, g^{\zeta \beta} \gamma^{\rho} \gamma^{\alpha} E_{\alpha \beta} \Bigr) - \nonumber \\[3pt]
    && \qquad \qquad \qquad \qquad \;\; \epsilon^{\zeta \beta \sigma \rho} \partial_{\sigma} \Bigl( \gamma_{\zeta} \gamma^{\alpha} G_{\alpha \beta} \Bigr) = 0 \, , \\[5pt] \label{eom2}
    && \partial_{\zeta} \Bigl( \sqrt{-g}\, g^{\rho \beta} \gamma^{\zeta} \gamma^{\alpha} G_{\alpha \beta} - \sqrt{-g}\, g^{\zeta \beta} \gamma^{\rho} \gamma^{\alpha} G_{\alpha \beta} \Bigr) + \nonumber \\[3pt]
    && \qquad \qquad \qquad \qquad \;\; \epsilon^{\zeta \beta \sigma \rho} \partial_{\sigma} \Bigl( \gamma_{\zeta} \gamma^{\alpha} E_{\alpha \beta} \Bigr) = 0 \, ,
\end{eqnarray}
where we introduced $\gamma^{\alpha} \equiv n^{\alpha'}\! e^{\alpha}_{\alpha'}$, and used the following identities for the Riemann tensor $R_{\alpha \beta \zeta \delta}$ and the vierbein $e^{\alpha}_{\alpha'}$:
\vspace*{-3pt}
\begin{eqnarray}
    && R_{\alpha \beta \zeta \delta} + R_{\alpha \delta \beta \zeta} + R_{\alpha \zeta \delta \beta} = 0 \, , \\[3pt]
    && D_{\mu} e^{\alpha}_{\alpha'} = 0 \, .
\end{eqnarray}
\begingroup
\allowdisplaybreaks

Eq.~\eqref{eom2} can be solved to obtain $\gamma^{\alpha} G_{\alpha \beta}$:
\begin{eqnarray}\label{sol2}
    &&\gamma^{\alpha} G_{\alpha \beta} = \frac{1}{2} \varepsilon^{\zeta \sigma \mu \nu} E_{\mu \nu} \gamma_{\zeta} g_{\sigma \beta} + P_{\beta} \, , \\[3pt] \label{pprop}
    && \gamma^{\alpha} P_{\alpha} = D_{\alpha} P^{\alpha} = 0 \, , \;\; \gamma^{\alpha} D_{\alpha} P^{\beta} = 0 \, .
\end{eqnarray}
In particular, the solution~\eqref{sol2} can be found by taking advantage of the identity
\begin{eqnarray}\label{kind}
    && K_{\alpha \beta} = \gamma_{\alpha} \gamma^{\mu} K_{\mu \beta} - \gamma_{\beta} \gamma^{\mu} K_{\mu \alpha} - \varepsilon_{\alpha \beta \gamma \sigma} \gamma^{\sigma} D^{\gamma} \, , \;\;\; \\[3pt]
    && \text{where} \quad D^{\gamma} \equiv \frac{1}{2} \varepsilon^{\rho \gamma \sigma \zeta} \gamma_{\rho} K_{\sigma \zeta} \, ,
\end{eqnarray}
which holds for any antisymmetric tensor $K_{\alpha \beta}$. Indeed, applying this identity to the tensor $K_{\alpha \beta} = \varepsilon^{\zeta \sigma \mu \nu} E_{\mu \nu} g_{\alpha \zeta} g_{\sigma \beta}/2$, and using the regularity of the four-potential $A_{\mu}$:
\begin{eqnarray}\label{regularA}
    \epsilon^{\mu \nu \lambda \rho}\, \partial_{\nu} E_{\lambda \rho} = 0 \, ,
\end{eqnarray}
one can transform Eq.~\eqref{eom2} into:
\begin{eqnarray}
    && D_{\zeta} \Bigl( L^{\rho} \gamma^{\zeta} - L^{\zeta} \gamma^{\rho} \Bigr) = 0 \, , \\[3pt] 
    && \text{where} \quad L_{\beta} \equiv \gamma^{\alpha} G_{\alpha \beta} - \frac{1}{2} \varepsilon^{\zeta \sigma \mu \nu} E_{\mu \nu} \gamma_{\zeta} g_{\sigma \beta} \, , \;\;\;
\end{eqnarray}
from which the solution~\eqref{sol2} trivially follows.

Now, we can substitute the solution~\eqref{sol2} into the Eq.~\eqref{eom1} to obtain:
\begin{eqnarray}\label{Aproppre}
\partial_{\mu} \left( \sqrt{-g}\, g^{\mu \lambda} g^{\nu \rho} \bar{E}_{\lambda \rho} \right) = 0 \, ,
\end{eqnarray}
where we used the identity~\eqref{kind} once again, and denoted $\bar{E}_{\lambda \rho} = E_{\lambda \rho} - \varepsilon_{\lambda \rho \mu \nu} \gamma^{\mu} P^{\nu}$. Due to Eqs.~\eqref{pprop}, $P_{\alpha}$ is a vector field defined on the 3-surfaces spanned locally by the vectors orthogonal to $\gamma^{\alpha}$, moreover this vector field is covariantly constant along the integral curves of $\gamma^{\alpha}$. Given Eq.~\eqref{sol2}, this means that $P_{\alpha}$ is fully determined by the Cauchy data for the EM field $E_{\alpha \beta}$ and $G_{\alpha \beta}$. It is therefore always possible to choose $P_{\beta} = 0$, that is $\bar{E}_{\lambda \rho} = E_{\lambda \rho}$. We are left with the following local-Lorentz-covariant equation for the four-potential $A_{\mu}$:
\begin{eqnarray}\label{Aprop}
\partial_{\mu} \left( \sqrt{-g}\, g^{\mu \lambda} g^{\nu \rho} {E}_{\lambda \rho} \right) = 0 \, .
\end{eqnarray}

Analogously to what was described in the previous paragraphs, one can first solve the Eq.~\eqref{eom1} to find $\gamma^{\alpha} E_{\alpha \beta}$, and then substitute the solution into the Eq.~\eqref{eom2}. In this case, one obtains the equation for the $C_{\mu}$ four-potential only:
\begin{eqnarray}\label{Cprop}
    \partial_{\mu} \left( \sqrt{-g}\, g^{\mu \lambda} g^{\nu \rho} G_{\lambda \rho} \right) = 0 \, .
\end{eqnarray}

We see that in the case of free electromagnetic field in curved spacetime, the equations for the two four-potentials $A_{\mu}$ and $C_{\mu}$ are identical. This is a manifestation of the symmetry of the Lagrangian~\eqref{Lgr}, which is invariant with respect to the $SO(2)$ rotation of the four-potentials:
\begin{equation}
\begin{aligned}\label{rot2}
    &{A_{\mu}} \; \rightarrow \; {A_{\mu}} \cos \theta - {C_{\mu}} \sin \theta \\
    &{C_{\mu}} \; \rightarrow \; {C_{\mu}} \cos \theta + {A_{\mu}} \sin \theta \, \, .
\end{aligned}
\end{equation}
The latter invariance gives rise to the following conserved Noether current:
\begin{equation}\label{helgen}
    s^{\nu} = \sqrt{-g}\, g^{\mu \lambda} g^{\nu \rho} \left( A_{\lambda} G_{\mu \rho} - C_{\lambda } E_{\mu \rho} \right) \, .
\end{equation}
As it will become evident after we introduce charged particles, this is exactly the curved spacetime generalization of the helicity four-current discussed in the previous section. One can now see that within the theory~\eqref{Lgr}, EOM for the EM field in the presence of gravity are fully consistent with the conservation of helicity:
\begin{equation}\label{emduin}
   D_{\mu} E^{\mu \nu} = 0, \quad  D_{\mu} G^{\mu \nu} = 0 \quad \Longleftrightarrow \quad D_{\mu} s^{\mu} = 0 \, .
\end{equation}

Finally, let us note that the fixed four-vector $n^{\alpha'}$ introduced in Eq.~\eqref{Lgr} does not enter the EOM~\eqref{Aprop} and~\eqref{Cprop} describing the propagation of the EM field: the dynamics of the EM field is fully local-Lorentz-invariant. It is also important that although we introduced the two four-potentials for the description of the EM field instead of the only one normally used, the number of degrees of freedom of the EM field remained the same. Similarly to the case of the flat spacetime Maxwell equations, the curved spacetime Eqs.~\eqref{eom1} and~\eqref{eom2} contain $6$ independent unknown functions, which are $\gamma^{\alpha} E_{\alpha \beta}$ and $\gamma^{\alpha} G_{\alpha \beta}$ (note the antisymmetricity of $E_{\alpha \beta}$ and $G_{\alpha \beta}$), and $6$ independent equations (note the divergencelessness of the left-hand-sides of Eqs.~\eqref{eom1} and~\eqref{eom2}).

\textit{Motion of charged particles.}---Let us now introduce charges into the theory. In particular, we consider the following Lagrangian density for the interactions between charged particles and the EM field:
\begin{eqnarray}\label{lint}
    \mathcal{L}_2 = \sqrt{-g}\, j_{\text{e}}^{\alpha} A_{\alpha} + \sqrt{-g}\, j_{\text{m}}^{\alpha} C_{\alpha} \, ,
\end{eqnarray}
where $j_{\text{e}}^{\alpha}$ and $j_{\text{m}}^{\alpha}$ are the electric and magnetic four-currents, respectively. In particular, the currents are given by the following expressions:
	\begin{eqnarray}
	j_{\text{e}}^{\nu} (x) = \sum_i q_i \int \frac{\delta^4 (x-x_i(\tau_i))}{\sqrt{-g}}\, dx_i^{\nu} \, , \\[3pt]
	j_{\text{m}}^{\nu} (x) = \sum_i g_i \int \frac{\delta^4 (x-x_i(\tau_i))}{\sqrt{-g}}\, dx_i^{\nu} \, , 
	\end{eqnarray}
where $x^{\nu}_i(\tau_i)$ is the worldline of the $i$-th particle, $q_i$ ($g_i$) is the electric (magnetic) charge of the $i$-th particle, $\delta^4 (x)$ is the Dirac delta distribution density. Note that from Eq.~\eqref{lint}, it follows that the $SO(2)$ rotation of the four-potentials~\eqref{rot2} corresponds to the rotation in the electric-magnetic plane, so that the four-current Eq.~\eqref{helgen} is indeed the helicity of the EM field. The EOM for the four-potentials $A_{\mu}$ and $C_{\mu}$ derived by varying the action $S= \int d^4x \left( \mathcal{L}_1 + \mathcal{L}_2 \right)$ are:
\begin{eqnarray}\label{meom1}
    && \partial_{\zeta} \Bigl( \sqrt{-g}\, g^{\rho \beta} \gamma^{\zeta} \gamma^{\alpha} E_{\alpha \beta} - \sqrt{-g}\, g^{\zeta \beta} \gamma^{\rho} \gamma^{\alpha} E_{\alpha \beta} \Bigr) - \nonumber \\
    && \qquad \qquad \;\;\; \epsilon^{\zeta \beta \sigma \rho} \partial_{\sigma} \Bigl( \gamma_{\zeta} \gamma^{\alpha} G_{\alpha \beta} \Bigr) = -\sqrt{-g}\, j_{\text{e}}^{\rho} \, , \\ \label{meom2}
    && \partial_{\zeta} \Bigl( \sqrt{-g}\, g^{\rho \beta} \gamma^{\zeta} \gamma^{\alpha} G_{\alpha \beta} - \sqrt{-g}\, g^{\zeta \beta} \gamma^{\rho} \gamma^{\alpha} G_{\alpha \beta} \Bigr) + \nonumber \\
    && \qquad \qquad \;\;\; \epsilon^{\zeta \beta \sigma \rho} \partial_{\sigma} \Bigl( \gamma_{\zeta} \gamma^{\alpha} E_{\alpha \beta} \Bigr) = -\sqrt{-g}\, j_{\text{m}}^{\rho} \, .
\end{eqnarray}

As in the previous section, we can transform Eq.~\eqref{meom2} as follows:
\begin{eqnarray}
    && \!\!\!\!\!\!\! D_{\zeta} \Bigl( L^{\rho} \gamma^{\zeta} - L^{\zeta} \gamma^{\rho} \Bigr) = -j^{\rho}_{\text{m}} \, , \\[3pt] 
    && \!\!\!\!\!\!\! \text{where} \quad L_{\beta} \equiv \gamma^{\alpha} G_{\alpha \beta} - \frac{1}{2} \varepsilon^{\zeta \sigma \mu \nu} E_{\mu \nu} \gamma_{\zeta} g_{\sigma \beta} \, ,
\end{eqnarray}
The solution for $\gamma^{\alpha} G_{\alpha \beta}$ is:
\begin{eqnarray}\label{sol4}
    \!\!\!\!\!\!\!\!\!\!\!\! \gamma^{\alpha} G_{\alpha \beta} = \frac{1}{2} \varepsilon^{\zeta \sigma \mu \nu} E_{\mu \nu} \gamma_{\zeta} g_{\sigma \beta} - g_{\alpha \beta} R j^{\alpha}_{\text{m}} + P_{\beta} \, ,
\end{eqnarray}
where $P_{\beta}$ has the properties detailed in Eqs.~\eqref{pprop}, so one can always impose $P_{\beta}=0$ as discussed in the previous section; $R \equiv \left( \gamma^{\mu} D_{\mu} \right)^{-1}$ is an integral operator with the following kernel:
\begin{eqnarray}\label{rk}
    R_K (x-y) = \int_0^{\infty}\! \delta^4 (x-y-\gamma s)\,  ds \, .
\end{eqnarray}
We then substitute the solution~\eqref{sol4} into Eq.~\eqref{meom1} to obtain:
\begin{eqnarray}\label{Aeq}
\partial_{\mu} \left( \sqrt{-g}\, g^{\mu \lambda} g^{\nu \rho} E'_{\lambda \rho} \right) = -\sqrt{-g}\, j_{\text{e}}^{\nu} \, ,
\end{eqnarray}
where
\begin{eqnarray}\label{Rjm}
    E'_{\lambda \rho} = E_{\lambda \rho} + \varepsilon_{\lambda \rho \mu \nu} \gamma^{\mu} R j^{\nu}_{\text{m}} \, .
\end{eqnarray}
Similarly, we find that Eq.~\eqref{meom1} admits the following solution:
\begin{eqnarray}\label{sol3}
    \!\!\!\!\!\!\!\!\!\!\!\! \gamma^{\alpha} E_{\alpha \beta} = -\frac{1}{2} \varepsilon^{\zeta \sigma \mu \nu} G_{\mu \nu} \gamma_{\zeta} g_{\sigma \beta} - g_{\alpha \beta} R j^{\alpha}_{\text{e}} \, .
\end{eqnarray}
Substituting it into Eq.~\eqref{meom2}, we obtain the following equation for the $C_{\mu}$ four-potential:
\begin{eqnarray}\label{Ceq}
\partial_{\mu} \left( \sqrt{-g}\, g^{\mu \lambda} g^{\nu \rho} G'_{\lambda \rho} \right) = -\sqrt{-g}\, j_{\text{m}}^{\nu} \, ,
\end{eqnarray}
where
\begin{eqnarray}\label{Rje}
    G'_{\lambda \rho} = G_{\lambda \rho} - \varepsilon_{\lambda \rho \mu \nu} \gamma^{\mu} R j^{\nu}_{\text{e}} \, .
\end{eqnarray}
Eqs.~\eqref{Rjm} and~\eqref{Rje} can be used to reformulate the conditions of the regularity of the four-potentials (i.e. Eq.~\eqref{regularA} and its analogue $\epsilon^{\mu \nu \lambda \rho} \partial_{\nu} G_{\lambda \rho}=0$), in terms of the tensors $E'_{\mu\nu}$ and $G'_{\mu\nu}$:
\begin{eqnarray}\label{reg1}
    && \frac{1}{2}\, \epsilon^{\mu \nu \lambda \rho} \partial_{\nu} E'_{\lambda \rho} = \sqrt{-g}\, j_{\text{m}}^{\mu} \, , \\[3pt] \label{reg2}
    && \frac{1}{2}\, \epsilon^{\mu \nu \lambda \rho} \partial_{\nu} G'_{\lambda \rho} = -\sqrt{-g}\, j_{\text{e}}^{\mu} \, ,
\end{eqnarray}
where we used the conservation of the charged currents $D_{\mu} j_{\text{e}}^{\mu} = D_{\mu} j_{\text{m}}^{\mu} = 0$. Comparing these equations with Eqs.~\eqref{Aeq} and~\eqref{Ceq}, one sees that it is always possible to impose the following condition:
\begin{eqnarray}\label{mainrel}
    \frac{1}{2}\, \epsilon^{\mu \nu \lambda \rho} E'_{\lambda \rho} = \sqrt{-g}\, g^{\nu \lambda} g^{\mu \rho} G'_{\lambda \rho} \, .
\end{eqnarray}

From Eqs.~\eqref{Aeq} and~\eqref{Ceq}, we see that charged particles source the field described by the tensors $E'_{\mu\nu}$ and $G'_{\mu\nu}$ in a fully local-Lorentz-invariant way. However, Eqs.~\eqref{Rjm} and~\eqref{Rje} show that the latter field is related to the tensors $E_{\mu \nu}$ and $G_{\mu \nu}$ in a way that breaks local Lorentz invariance: the difference depends on a fixed Lorentz four-vector $n^{\alpha'}$. It is a very interesting observation made by Zwanziger in Ref.~\cite{Zwanziger:1968rs}, that the flat spacetime analogues of the would-be Lorentz-violating terms in Eqs.~\eqref{Rjm} and~\eqref{Rje} have no influence on physical observables: while the theory is not manifestly Lorentz-invariant, all the possible observable processes described by it are Lorentz-invariant. Let us now show that the curved spacetime case is similar in this regard: all the physics described by the Lagrangian densities~\eqref{Lgr} and~\eqref{lint} is local-Lorentz-invariant.

For this, we have to consider the EOM of charged particles in an EM field. While the left-hand side of these EOM is given by the standard expression, the right-hand side can be easily derived by extremizing the action with the Lagrangian density~\eqref{lint}. The result is:
\begin{eqnarray}\label{geod}
    g_{\lambda \rho} m_i u_i^{\nu} D_{\nu} u_i^{\rho} = \Bigl( q_i E_{\lambda \mu} (x_i) + g_i G_{\lambda \mu} (x_i) \Bigr) u_i^{\mu} \, ,
\end{eqnarray}
where $m_i$ is the mass of the $i$-th particle and $u_i^{\nu} = dx_i^{\nu}/d\tau_i$. Let us now rewrite these EOM in terms of the local-Lorentz-covariant tensors $E'_{\lambda \rho}$ and $G'_{\lambda \rho}$ entering Eqs.~\eqref{Aeq} and~\eqref{Ceq}:
\begin{eqnarray}\label{llv}
    && g_{\lambda \rho} m_i u_i^{\nu} D_{\nu} u_i^{\rho} = \Bigl( q_i E'_{\lambda \mu} (x_i) + g_i G'_{\lambda \mu} (x_i) \Bigr) u_i^{\mu} + \nonumber \\[3pt]
	&& \sum_j (q_i g_j - g_i q_j) \!\! \int \!\! R_K (x_i - x_j)\, \epsilon_{\mu \lambda \nu \rho} \gamma^{\mu} u^{\nu}_i u^{\rho}_j\, d\tau_j \, . \quad \;\;
\end{eqnarray}
Due to the form of the kernel $R_K$ (Eq.~\eqref{rk}), the local-Lorentz-violating term in Eq.~\eqref{llv} is non-zero only if the worldline of the $i$-th particle satisfies:
\begin{eqnarray}\label{nodense}
    x^{\nu}_i (\tau_i) = x^{\nu}_j (\tau_j) + \gamma^{\nu} s \, ,
\end{eqnarray}
for any $\tau_i \, , \tau_j \, , s \in (-\infty , \infty)$. Since there are more equations than free parameters in this condition, it is satisfied only for exceptional worldlines, i.e. the set of all the worldlines $x_i(\tau_i)$ satisfying Eq.~\eqref{nodense} is nowhere dense in spacetime. An interesting feature of nowhere dense sets of worldlines is that they can never be measured, as every measurement has a finite accuracy. Thus, the last term on the right-hand side of Eq.~\eqref{llv} can be safely omitted, and we are left with the same form for the EOM as in Eq.~\eqref{geod}, but now with the EM field represented by the local-Lorentz-covariant tensors $E'_{\lambda \mu}$ and $G'_{\lambda \mu}$. We therefore see that the classical dynamics described by the theory with the Lagrangian density $\mathcal{L}_1 + \mathcal{L}_2$ is fully local-Lorentz-invariant. 

One can wonder whether the local Lorentz invariance survives also at the quantum level. The answer is yes, if the Dirac-Schwinger-Zwanziger (DSZ) quantization condition is satisfied, in the full analogy to the flat spacetime case. Indeed, it is enough to show that for closed particle paths, the path integral does not depend on $n^{\alpha'}$~\cite{Brandt:1977be, Brandt:1978wc}. This means that the following action
\begin{eqnarray}
    \int d^4 x\, \mathcal{L}_2 = \int_{\Sigma_i}\!\! d\Sigma^{\alpha \beta} \Bigl( q_i E_{\alpha \beta} + g_i G_{\alpha \beta} \Bigr) \, ,
\end{eqnarray}
where $\Sigma_i$ is a 2-surface with the boundary $x_i (\tau_i)$, has to be $n^{\alpha'}$-independent modulo $2\pi n$ where $n\in \mathbb{Z}$. Using Eqs.~\eqref{Rjm} and~\eqref{Rje}, we find the $n^{\alpha'}$-dependent part of this action:
\begin{eqnarray}
    && \sum_j (q_i g_j - g_i q_j) \!\! \int_{\Sigma_i}\!\! d\Sigma^{\alpha \beta} \gamma^{\mu} \epsilon_{\alpha \beta \mu \nu} \times \nonumber \\[3pt]
    && \oint dx_j^{\nu} \int_0^{\infty} \!\!\! ds\; \delta^4 (x-x_j-\gamma s) = (q_i g_j - g_i q_j) L_{ij} \, , \quad \;\;\;\;
\end{eqnarray}
where $L_{ij} \in \mathbb{Z}$ is the Gauss linking number counting the number of times the loop $x_j (\tau)$ intersects the oriented 3-surface $\Sigma_i\! \times \! \gamma s$, $s\in [ 0, \infty)$. One sees that the path integral of the theory is $n^{\alpha'}$-independent as long as the DSZ quantization condition is satisfied: $q_i g_j - g_i q_j = 2\pi n$, $n\in \mathbb{Z}$.

\textit{Electric and magnetic fields.}---In the previous sections, we introduced the theory of the EM field in curved spacetime using a general-covariant formalism. We showed that the latter theory is electric-magnetic duality invariant, see Eq.~\eqref{emduin}. Still, we started our discussion of the electric-magnetic duality symmetry by writing down the curved spacetime Maxwell equations~\eqref{nomm}--\eqref{mmlas} in terms of electric and magnetic fields and noticing the asymmetry between $\mathbf{E}$ and $\mathbf{B}$. Let us see how the EM field equations~\eqref{Aeq} and~\eqref{Ceq} written in terms of electric and magnetic fields are reconciled with the electric-magnetic duality symmetry.

The key observation is that the EM field need not be defined in terms of the only one pair ($\mathbf{E}$, $\mathbf{B}$) of electric and magnetic fields. Indeed, the relevant dynamical variables for the EM field are four-potentials, while the description in terms of electric and magnetic fields is a matter of convenience. For example, whenever our system admits the presence of hypothetical magnetic monopoles, there are two different ways to define the magnetic field: it can either be defined as a force acting on a probe magnetic charge at rest $\mathbf{B}_{\text{m}}$, or it can be defined in a standard way via a force acting on an electric current $\mathbf{B}_{\text{e}}$. It turns out that in a general curved spacetime background, these two definitions give different results $\mathbf{B}_{\text{e}} \neq \mathbf{B}_{\text{m}}$. In particular, guided by the EOM for the charges Eq.~\eqref{llv}, we define the electric and magnetic fields as follows:
\begin{eqnarray}\label{eebm}
&&E_{\text{e}}{}_i = -E'_{0i}\, , \;\; B_{\text{m}}{}_i = -G'_{0i} \, , \\  \label{beem}
&&B_{\text{e}}{}_i = \frac{1}{2}\, \epsilon_{ijk} E'_{jk}\, , \;\; E_{\text{m}}{}_i = -\frac{1}{2}\, \epsilon_{ijk} G'_{jk}\, ,
\end{eqnarray}
so that $\mathbf{E}_{\text{e}}$ and $\mathbf{B}_{\text{e}}$ determine the Lorentz force acting on the electric charges, while $\mathbf{B}_{\text{m}}$ and $\mathbf{E}_{\text{m}}$ determine the dual Lorentz force acting on the magnetic charges.
Then, the EM field equations in curved spacetime~\eqref{Aeq} and~\eqref{Ceq}  become:
\begingroup
\allowdisplaybreaks
\begin{eqnarray}\label{maxtwo1}
&& \pmb{\nabla}\! \cdot \! \left\lbrace \sqrt{-g} \left[ g^{00} \mathcal{G}\!\cdot \! \mathbf{E}_{\text{e}} - \mathbf{g} \left( \mathbf{g}\! \cdot \! \mathbf{E}_{\text{e}} \right) + \mathcal{G} \! \cdot \! \left( \mathbf{g}\! \times \! \mathbf{B}_{\text{e}} \right) \right] \right\rbrace = \nonumber \\
&& \qquad \qquad \qquad \qquad \qquad \qquad \qquad \qquad \; - \sqrt{-g} \rho_{\text{e}} \, , \\[3pt]
&& \pmb{\nabla}\! \times \! \left\lbrace \sqrt{-g} \left[ \mathbf{g} \!\times \! \left( \mathcal{G}\! \cdot \! \mathbf{E}_{\text{e}} \right)  + \frac{1}{2}\, \mathrm{tr} \left( \mathcal{G} \mathbf{S} \mathcal{G} \mathbf{S} \! \cdot \! \mathbf{B}_{\text{e}} \right)  \right] \right\rbrace - \nonumber \\
&& \; \frac{\partial}{\partial t} \left\lbrace \sqrt{-g} \left[ g^{00} \mathcal{G}\!\cdot \! \mathbf{E}_{\text{e}} - \mathbf{g} \left( \mathbf{g}\! \cdot \! \mathbf{E}_{\text{e}} \right) + \mathcal{G} \! \cdot \! \left( \mathbf{g}\! \times \! \mathbf{B}_{\text{e}} \right) \right] \right\rbrace = \nonumber \\
&& \qquad \qquad \qquad \qquad \qquad \qquad \qquad \qquad \; - \sqrt{-g}\, \mathbf{j}_{\text{e}} \, , \\[3pt]
&& \pmb{\nabla}\! \cdot \! \left\lbrace \sqrt{-g} \left[ g^{00} \mathcal{G}\!\cdot \! \mathbf{B}_{\text{m}} - \mathbf{g} \left( \mathbf{g}\! \cdot \! \mathbf{B}_{\text{m}} \right) - \mathcal{G} \! \cdot \! \left( \mathbf{g}\! \times \! \mathbf{E}_{\text{m}} \right) \right] \right\rbrace = \nonumber \\
&& \qquad \qquad \qquad \qquad \qquad \qquad \qquad \qquad - \sqrt{-g} \rho_{\text{m}} \, , \\[3pt]
&& \pmb{\nabla}\! \times \! \left\lbrace \sqrt{-g} \left[ \mathbf{g} \!\times \! \left( \mathcal{G}\! \cdot \! \mathbf{B}_{\text{m}} \right)  - \frac{1}{2}\, \mathrm{tr} \left( \mathcal{G} \mathbf{S} \mathcal{G} \mathbf{S} \! \cdot \! \mathbf{E}_{\text{m}} \right)  \right] \right\rbrace - \nonumber \\
&& \; \frac{\partial}{\partial t} \left\lbrace \sqrt{-g} \left[ g^{00} \mathcal{G}\!\cdot \! \mathbf{B}_{\text{m}} - \mathbf{g} \left( \mathbf{g}\! \cdot \! \mathbf{B}_{\text{m}} \right) - \mathcal{G} \! \cdot \! \left( \mathbf{g}\! \times \! \mathbf{E}_{\text{m}} \right) \right] \right\rbrace = \nonumber \\
&& \qquad \qquad \qquad \qquad \qquad \qquad \qquad \qquad - \sqrt{-g}\, \mathbf{j}_{\text{m}} \, ,
\end{eqnarray}
\endgroup
where the notation from Eqs.~\eqref{nomm}--\eqref{mmlas} is used, and the components of the four-currents are introduced $j^{\mu}_i = (\rho_i, \mathbf{j}_i)$.
These equations can be complemented by the conditions of the regularity of the four-potentials Eqs.~\eqref{reg1} and~\eqref{reg2}:
\begin{eqnarray}
&&\pmb{\nabla}\! \cdot \! \mathbf{B}_{\text{e}} = \sqrt{-g}\,\rho_{\text{m}} \, , \\[3pt] 
&& \pmb{\nabla}\! \times \! \mathbf{E}_{\text{e}} + \frac{\partial  \mathbf{B}_{\text{e}} }{\partial t} = - \sqrt{-g}\,\mathbf{j}_{\text{m}} \, , \\[3pt]
&&\pmb{\nabla}\! \cdot \! \mathbf{E}_{\text{m}} = \sqrt{-g}\, \rho_{\text{e}} \, , \\[3pt] \label{maxtwol}
&& \pmb{\nabla}\! \times \! \mathbf{B}_{\text{m}} - \frac{\partial  \mathbf{E}_{\text{m}} }{\partial t} = \sqrt{-g}\,\mathbf{j}_{\text{e}} \, .
\end{eqnarray}
The relations between different electric and magnetic fields Eqs.~\eqref{mainrel} in the 3-vector form are:
\vspace*{-2pt}
\begin{eqnarray}
   &&\!\!\!\!\!\!\!\!\! \mathbf{E}_{\text{e}} = \sqrt{-g} \left[ \mathbf{g} \!\times \! \left( \mathcal{G}\! \cdot \! \mathbf{B}_{\text{m}} \right)  - \frac{1}{2}\, \mathrm{tr} \left( \mathcal{G} \mathbf{S} \mathcal{G} \mathbf{S} \! \cdot \! \mathbf{E}_{\text{m}} \right)  \right] \, , \\[3pt]
   &&\!\!\!\!\!\!\!\!\! \mathbf{B}_{\text{e}} = - \sqrt{-g} \left[ g^{00} \mathcal{G}\!\cdot \! \mathbf{B}_{\text{m}} - \mathbf{g} \left( \mathbf{g}\! \cdot \! \mathbf{B}_{\text{m}} \right) - \mathcal{G} \! \cdot \! \left( \mathbf{g}\! \times \! \mathbf{E}_{\text{m}} \right) \right] \, , \qquad
   \vspace*{-2pt}
\end{eqnarray}
or equivalently:
\vspace*{-2pt}
\begin{eqnarray}
   &&\!\!\!\!\!\!\!\!\!\!\! \mathbf{B}_{\text{m}} = -\sqrt{-g} \left[ \mathbf{g} \!\times \! \left( \mathcal{G}\! \cdot \! \mathbf{E}_{\text{e}} \right)  + \frac{1}{2}\, \mathrm{tr} \left( \mathcal{G} \mathbf{S} \mathcal{G} \mathbf{S} \! \cdot \! \mathbf{B}_{\text{e}} \right)  \right] \, , \\[3pt]
   &&\!\!\!\!\!\!\!\!\!\!\! \mathbf{E}_{\text{m}} = -\sqrt{-g} \left[ g^{00} \mathcal{G}\!\cdot \! \mathbf{E}_{\text{e}} - \mathbf{g} \left( \mathbf{g}\! \cdot \! \mathbf{E}_{\text{e}} \right) + \mathcal{G} \! \cdot \! \left( \mathbf{g}\! \times \! \mathbf{B}_{\text{e}} \right) \right] \, . \qquad 
   \vspace*{-2pt}
\end{eqnarray}

The system of Eqs.~\eqref{maxtwo1}--\eqref{maxtwol} is symmetric with respect to the $SO(2)$ electric-magnetic duality transformations of the electric and magnetic fields supplemented by the 
interchange $\sigma$ of the indices $\text{e}$ and $\text{m}$:
\vspace*{-2pt}
\begin{equation}
\begin{aligned}\label{rotnew}
    &\mathbf{E}_{\alpha} \; \rightarrow \; \mathbf{E}_{\sigma (\alpha )} \cos \theta + \mathbf{B}_{\sigma (\alpha )} \sin \theta \\
    &\mathbf{B}_{\alpha} \; \rightarrow \; \mathbf{B}_{\sigma (\alpha )}  \cos \theta - \mathbf{E}_{\sigma (\alpha )}  \sin \theta \, \, ,
    \vspace*{-3pt}
\end{aligned}
\end{equation}
as well as by the similar $SO(2)$ rotation of the charges and currents.

Since there are only electrically charged particles in the laboratory, the part of the EM field EOM~\eqref{maxtwo1}--\eqref{maxtwol} relevant for the GW detection features only the $\mathbf{E}_e$ and $\mathbf{B}_e$ fields and therefore has the same structure as Eqs.~\eqref{nomm}--\eqref{mmlas}. This means that the duality-symmetric theory of the EM field in curved spacetime presented here yields no novel predictions for the GW detection experiments.

\textit{Conclusions.}---Ever since the start of the XXth century, there has been a prevailing point of view in physics community that the electromagnetic field represents a physical entity separate from charged particles used to measure it. In this article, we showed that this perspective, which is currently adopted in any exhaustive quantization scheme, leads to the tension within the standard description of the electromagnetic field in General Relativity, since the core principle of General Relativity -- the equivalence principle -- as applied to the electromagnetic field itself, is violated. 
In particular, we showed that in the standard description, the otherwise conserved Noether charge corresponding to the electric-magnetic duality symmetry of the electromagnetic field is no longer conserved in the presence of gravity. We also showed that even if one rejects the above-mentioned notion of the electromagnetic field as an entity independent of the charges used to probe it, the standard description still conflicts with the equivalence principle whenever hypothetical magnetic charges are allowed to exist. Meanwhile, the possibility that magnetic charges exist is considered highly likely within the theoretical physics community due to the phenomenon of the quantization of charge observed in nature~\cite{Dirac:1931kp}, due to the predictions of Grand Unified Theories~\cite{tHooft:1974kcl, Polyakov:1974ek}, as well as due to the implications of the general requirement that quantum theory and gravity coexist~\cite{Banks:2010zn}.

Motivated by these tensions within the standard description, we constructed an electric-magnetic duality symmetric theory of the interactions between electromagnetic field and gravity. In particular, we proposed a general-covariant generalization of the flat spacetime two-potential formalism developed earlier in Ref.~\cite{Zwanziger:1968rs}. We showed that although the Lagrangian of the theory is not manifestly local-Lorentz-invariant, all the physics described by it respects local Lorentz symmetry. We also found an interesting difference with respect to the flat spacetime case: in general, the electromagnetic field in the presence of gravity has to be described in terms of two pairs of electric and magnetic fields. While the one pair of fields can be defined by their action on a probe electric charge or current, the other pair can be defined by their action on a probe magnetic charge or current. The duality symmetry transformation then necessarily involves an interchange of these two pairs of fields.

\begin{acknowledgments}
\textit{Acknowledgments.}---
The author thanks an anonymous referee and Andreas Ringwald for valuable comments on the manuscript, as well as Kevin Zhou, Sebastian A.~R.~Ellis and Jai-chan Hwang for fruitful discussions. The author is funded by the UK Research and Innovation grant MR/V024566/1. During his time at DESY, the author was funded by the Deutsche Forschungsgemeinschaft (DFG, German Research Foundation) under Germany's Excellence Strategy -- EXC 2121 \textit{Quantum Universe} -- 390833306.
\vspace{-10pt}
\end{acknowledgments}
\endgroup
\vspace{-6pt}

\bibliography{emgravdual.bib}

\end{document}